\documentclass[longauth]{aa} 
\usepackage{graphicx}
\usepackage{tabularx}
\usepackage{multirow}
\usepackage[colorlinks=true, linkcolor=blue, citecolor=blue]{hyperref}
\usepackage{gensymb}
\usepackage{color}
\usepackage{txfonts}
\usepackage{natbib}
\usepackage{amsmath}
\usepackage{ulem}
\usepackage[flushleft]{threeparttable}
\usepackage[dvipsnames]{xcolor}
\usepackage{rotating}
\usepackage{CJKutf8}

\newcommand{\JWST}{\textit{JWST}}
\newcommand{\HST}{\textit{HST}}

\newcommand{\cii}{[C\,\textsc{ii}]}
\newcommand{\oiii}{[O\,\textsc{iii}]}

\newcommand{\oi}{[O\,\textsc{iii}]}

\newcommand{\fluxunit}{erg\,s$^{-1}$\,cm$^{-2}$}

\begin{document} 
   \title{No [CII] or dust detection in two Little Red Dots at $z_{\rm spec}>7$}

\author{Mengyuan Xiao\inst{\ref{inst1}}\thanks{E-mail: mengyuan.xiao@unige.ch}
\and Pascal A. Oesch\inst{\ref{inst1},\ref{inst2},\ref{inst21}}
\and Longji Bing\inst{\ref{inst3}}
\and David Elbaz\inst{\ref{inst4}}
\and Jorryt Matthee\inst{\ref{inst5}} 
\and Yoshinobu Fudamoto\inst{\ref{inst6},\ref{inst7}} 
\and Seiji Fujimoto\inst{\ref{inst8},\ref{inst9},\ref{inst10}}
\and Rui Marques-Chaves\inst{\ref{inst1}} 
\and Christina C. Williams\inst{\ref{inst11},\ref{inst7}} 
\and Miroslava Dessauges-Zavadsky\inst{\ref{inst1}} 
\and Francesco Valentino\inst{\ref{inst2},\ref{inst20}} 
\and Gabriel Brammer\inst{\ref{inst2},\ref{inst21}} 
\and Alba Covelo-Paz\inst{\ref{inst1}}
\and Emanuele Daddi\inst{\ref{inst4}} 
\and Johan P. U. Fynbo\inst{\ref{inst2},\ref{inst21}} 
\and Steven Gillman \inst{\ref{inst2},\ref{inst20}} 
\and Michele Ginolfi\inst{\ref{inst13},\ref{inst14}} 
\and Emma Giovinazzo\inst{\ref{inst1}}
\and Jenny E. Greene\inst{\ref{inst15}} 
\and Qiusheng Gu\inst{\ref{inst16},\ref{inst17}} 
\and Garth Illingworth\inst{\ref{inst24}} 
\and Kohei Inayoshi\inst{\ref{inst12}} 
\and Vasily Kokorev\inst{\ref{inst9}} 
\and Romain A. Meyer\inst{\ref{inst1}} 
\and Rohan P. Naidu\inst{\ref{inst18}, \ref{inst8}} 
\and Naveen A. Reddy\inst{\ref{inst22}}
\and Daniel Schaerer\inst{\ref{inst1}} 
\and Alice Shapley\inst{\ref{inst23}}
\and Mauro Stefanon\inst{\ref{inst25},\ref{inst26}} 
\and Charles L. Steinhardt\inst{\ref{inst27}} 
\and David J. Setton\inst{\ref{inst15}} 
\and Marianne Vestergaard\inst{\ref{inst19},\ref{inst7}} 
\and Tao Wang\inst{\ref{inst16},\ref{inst17}}   
\\
}

\institute{Department of Astronomy, University of Geneva, Chemin Pegasi 51, 1290 Versoix, Switzerland\label{inst1}
\and Cosmic Dawn Center (DAWN), Denmark \label{inst2}
\and  Niels Bohr Institute, University of Copenhagen, Jagtvej 128, K\o benhavn N, DK-2200, Denmark \label{inst21}
\and Astronomy Centre, University of Sussex, Falmer, Brighton BN1 9QH, UK  \label{inst3}
\and Universit{\'e} Paris-Saclay, Universit{\'e} Paris Cit{\'e}, CEA, CNRS, AIM, 91191, Gif-sur-Yvette, France \label{inst4}
\and Institute of Science and Technology Austria (ISTA), Am Campus 1, 3400 Klosterneuburg, Austria \label{inst5}
\and Center for Frontier Science, Chiba University, 1-33 Yayoi-cho, Inage-ku, Chiba 263-8522, Japan \label{inst6}
\and Steward Observatory, University of Arizona, 933 N Cherry Avenue, Tucson, AZ 85721, USA \label{inst7}
\and Hubble Fellow\label{inst8}
\and Department of Astronomy, The University of Texas at Austin, Austin, TX 78712, USA\label{inst9}
\and David A. Dunlap Department of Astronomy and Astrophysics, University of Toronto, 50 St. George Street, Toronto, Ontario, M5S 3H4, Canada\label{inst10}
\and NSF National Optical-Infrared Astronomy Research Laboratory, 950 North Cherry Avenue, Tucson, AZ 85719, USA \label{inst11}
\and DTU-Space, Elektrovej, Building 327 , 2800, Kgs. Lyngby, Denmark\label{inst20}
\and Dipartimento di Fisica e Astronomia, Università di Firenze, Via G. Sansone 1, I-50019, Sesto F.no (Firenze), Italy\label{inst13}
\and  INAF - Osservatorio Astrofisico di Arcetri, Largo E. Fermi 5, I-50125, Florence, Italy\label{inst14}
\and Department of Astrophysical Sciences, Princeton University, 4 Ivy Lane, Princeton, NJ 08544, USA\label{inst15}
\and School of Astronomy and Space Science, Nanjing University, Nanjing, Jiangsu 210093, China\label{inst16}
\and Key Laboratory of Modern Astronomy and Astrophysics, Nanjing University, Ministry of Education, Nanjing 210093, China\label{inst17}
\and  Department of Astronomy and Astrophysics, University of California, Santa Cruz, CA 95064, USA \label{inst24}
\and Kavli Institute for Astronomy and Astrophysics, Peking University, Beijing 100871, China \label{inst12}
\and MIT Kavli Institute for Astrophysics and Space Research, 70 Vassar Street, Cambridge, MA 02139, USA\label{inst18}
\and  Department of Physics and Astronomy, University of California, Riverside, 900 University Avenue, Riverside, CA 92521, USA \label{inst22}
\and  Department of Physics \& Astronomy, University of California, Los Angeles, 430 Portola Plaza, Los Angeles, CA 90095, USA \label{inst23}
\and  Departament d'Astronomia i Astrof\`isica, Universitat de Val\`encia, C. Dr. Moliner 50, E-46100 Burjassot, Val\`encia,  Spain\label{inst25}
\and  Unidad Asociada CSIC ``Grupo de Astrof\'isica Extragal\'actica y Cosmolog\'ia" (Instituto de F\'isica de Cantabria - Universitat de Val\`encia)\label{inst26}
\and Department of Physics and Astronomy, University of Missouri, 701 S. College Ave., Columbia, MO 65201, USA\label{inst27}
\and DARK, The Niels Bohr Institute, Jagtvej 155, 2200 Copenhagen N, Denmark\label{inst19}
}
\date{Received xxx; accepted xxx}

  \abstract
 {Little Red Dots (LRDs) are compact, point-like sources characterized by their red color and broad Balmer lines, which have been debated to be either dominated by active galactic nuclei (AGN) or dusty star-forming galaxies (DSFGs). Here we report two LRDs (ID9094 and ID2756) at $z_{\rm spec}>7$, recently discovered in the JWST FRESCO GOODS-North field. Both satisfy the ``v-shape" colors and compactness criteria for LRDs and are identified as Type-I AGN candidates based on their broad H$\beta$ emission lines (full width at half maximum: $2280\pm490$ km s$^{-1}$ for ID9094 and $1070\pm240$ km s$^{-1}$ for ID2756) and narrow \oi\, lines ($\simeq 300-400$ km s$^{-1}$). To investigate their nature, we conduct deep NOEMA follow-up observations targeting the \cii$\,158\,{\rm \mu m}$ emission line and the 1.3 mm dust continuum. We do not detect \cii\, or 1.3 mm continuum emission for either source. Notably, in the scenario that the two LRDs were DSFGs, we would expect significant detections: $>16\sigma$ for \cii\ and $>3\sigma$ for the 1.3 mm continuum of ID9094, and $>5\sigma$ for \cii\ of ID2756. Using the $3\sigma$ upper limits of \cii\, and 1.3 mm, we perform two analyses: (1) UV-to-FIR spectral energy distribution (SED) fitting with and without AGN components, and (2) comparison of their properties with the $L_{\rm \cii}$–SFR$_{\rm tot}$ empirical relation. Both analyses are consistent with a scenario where AGN activity may contribute to the observed properties, though a dusty star-forming origin cannot be fully ruled out. Our results highlight the importance of far-infrared observations for studying LRDs, a regime that remains largely unexplored.  }

   \maketitle
%

\section{Introduction}
Little Red Dots \citep[LRDs; e.g.,][]{Fujimoto2022Natur,Labbe2025, Matthee2024, Barro2024} have emerged as one of the most intriguing populations studied with the JWST \citep{Gardner2023}. The term "LRDs" was first introduced by \cite{Matthee2024} to describe galaxies exhibiting broad H$\alpha$ emission lines, with a red point-source morphology. In a broader context, the name LRD has also been applied to a variety of compact, red sources -- predominantly at high redshifts ($z\gtrsim4$) -- with ``v-shaped" (blue UV and red optical slopes in rest-frame) spectral energy distributions (SEDs), based mainly on photometric measurements only. The samples of these sources overlap, contributing to the complexity of their classification. Most recent studies agree that LRDs involve a combination of stellar and AGN contributions, although there is ongoing debate over which component dominates \citep[e.g.,][]{Wangbingjie2024, Akins2025}. Some works emphasize a dusty star-forming galaxy-dominated interpretation \citep[e.g.,][]{Perez-Gonzalez2024, Labbe2023a, Baggen2024, Williams2024_RED}, while others find stronger evidence for AGN dominance \citep[e.g.,][]{Kocevski2023, Kocevski2025, Greene2024, Kokorev2024, Ji2025, Naidu2025_bhstar,deGraaff2025_lrd}.

Multi-wavelength studies of LRDs have primarily focused on rest-frame UV, optical, near-infrared (NIR), mid-infrared (MIR), X-ray, and radio data, whereas far-infrared (FIR) studies remain relatively scarce in comparison. FIR observations, however, are crucial for disentangling the roles of AGN activity and dusty starburst processes, as they directly probe dust properties, star formation rates, and possible AGN contributions. Some studies have started to investigate LRDs in the FIR \citep[e.g.,][]{Labbe2025,Casey2024, Casey2025, Akins2024, Akins2025b, Williams2024_RED, Setton2025}, but most rely on photometric redshifts -- either through stacked analyses or individual source studies -- which might introduce large uncertainties in interpreting their dust properties and star formation activity. Current evidence shows that many LRDs are faint in X-rays \citep[e.g.,][]{Ananna2024,Maiolino2025,Yue2024, Sacchi2025}, exhibit weak or no radio emission \citep[e.g.,][]{Akins2024}, and lack detectable hot dust in the MIR \citep[e.g.,][]{Perez-Gonzalez2024,Williams2024_RED}. These properties could result from heavily obscured AGNs, extreme dusty starbursts, or even hybrid scenarios. Without robust FIR constraints, however, distinguishing between these possibilities remains challenging.

Another potential issue is the partial overlap in photometric selection methods for LRDs \citep[e.g.,][]{Labbe2025,Greene2024,Kokorev2024} and optical/NIR-selected DSFGs \citep[optically dark/faint galaxies; e.g.,][]{wangtao2019, Franco2018, Alcalde2019, Williams2019, Gomez-Guijarro2023, Xiao2023, Gomez-Guijarro2022a, Barrufet2023, Williams2024_RED, McKinney2023, Akins2023, Perez-Gonzalez2023_hstdark, Barro2024}, due to their similar red colors. This may pose a challenge for DSFG studies, as optical/NIR-selected DSFGs typically have more relaxed selection criteria, relying solely on color and/or magnitude cuts without considering morphology. As a result, LRDs could inadvertently be included in DSFG samples. Additionally, studies also show that some DSFGs have compact morphology \citep[e.g.,][]{Gomez-Guijarro2023}. This overlap introduces ambiguity, as both populations could appear in the same sample category despite potentially distinct physical origins. For example, DSFGs are typically powered by intense star formation obscured by large amounts of dust, while AGN-driven LRDs may exhibit similar photometric properties due to their reddened SEDs but have fundamentally different drivers of energy output.

In this study, we investigate two LRDs, ID9094 and ID2756\footnote{ID9094 and ID2756 correspond to gn17537 and gn28924, respectively, in the FRESCO \oiii\, catalog \citep{Meyer2024}.}, located in the JWST FRESCO GOODS-North field \citep{Oesch2023}. Due to the overlap in photometric selection criteria between DSFGs and LRDs, these two sources were initially classified as optically faint, dust-obscured galaxy candidates with red colors (F182M $-$ F444W $>$ 1.5 magnitude; hereafter mag) and as the most massive candidates at $z_{\rm spec} > 7$ ($M_{\star}> 10^{10} ~M_{\odot}$) based on optical-to-near-IR SED fitting \citep{Xiao2024}. However, as revealed in this study, deep JWST NIRCam grism spectra later reveal the presence of broad H$\beta$ emission lines (broad component of with full width at half maximum $>$ 1000 km s$^{-1}$; e.g., \cite{Matthee2024}; Sect.~\ref{Sec: lrd})\footnote{The broad lines were not clearly identified in the initial FRESCO spectra or at the time of the NOEMA proposal, due to the limited sensitivity of the FRESCO data.}, along with a red, point-source-like morphology, indicating potential AGN activity.

To further investigate their nature, we present follow-up FIR observations from NOEMA, targeting the \cii$\,158\,{\rm \mu m}$ emission line and the 1.3 mm dust continuum.  This study, together with the recent work by \cite{Setton2025}, represents one of the first efforts to investigate LRDs with both dust and \cii\, emission. \cite{Setton2025} focused on two of the most luminous LRDs at $z_{\rm spec} \sim 3-4.5$ and reported a deficit of hot and cold dust emission. In comparison, our study targets two LRDs at $z_{\rm spec} > 7$, a redshift regime more representative of the typical LRD population identified in JWST surveys. Together, the two studies provide a more complete view of the dust and gas properties of LRDs. Our NOEMA observations provide crucial constraints on the dust temperature, allowing us to assess whether the observed properties of these LRDs are consistent with DSFGs or if AGN-driven processes dominate. By combining FIR data with existing multi-wavelength photometry, this work addresses the critical question of whether DSFG scenarios can be ruled out for LRDs in the absence of spectroscopic confirmation.

This paper is organized as follows. In Section~\ref{Sec: data}, we describe the data and observational details. Section~\ref{Sec: result} presents our results, including the non-detection of the dust continuum and \cii\,line in the FIR and its implications for the nature of these LRDs. In Section~\ref{Sec: discussion}, we discuss the impact of LRDs on our understanding of DSFGs. We summarize our conclusions in Section~\ref{Sec: conclusion}.

Throughout this paper, we adopt a Chabrier initial mass function \citep[IMF;][]{Chabrier2003} to estimate SFR and stellar mass. We assume a Planck cosmology \citep{Planck2020} with $(\Omega_\mathrm{m},\, \Omega_\mathrm{\Lambda},\, h,\, \sigma_\mathrm{8})=(0.3,\, 0.7,\, 0.7, \,0.81)$. When necessary, data from the literature have been converted with a conversion factor of $M_{\star}$ \citep[][IMF]{Salpeter1955} = 1.7  $\times$ $M_{\star}$ \citep[][IMF]{Chabrier2003}. All magnitudes are in the AB system \citep{Oke1983}, such that $m_{\rm AB} = 23.9 - 2.5$ $\times$ log(S$_{\nu}$ [$\mu$Jy]).

\section{Data and sample}\label{Sec: data}
The two sources ID9094 ($\alpha$, $\delta$ [J2000] = 189.019240, 62.243531) at $z_{\rm spec}=7.0388\pm0.0001$ and ID2756 ($\alpha$, $\delta$ [J2000] = 189.083488,  62.202579) at $z_{\rm spec}=7.1883\pm0.0001$ are located in the GOODS-N field, and were first discovered in the JWST FRESCO survey \citep{Oesch2023} and reported in \cite{Xiao2024}.  They have red colors (F182M $-$ F444W $>$ 1.5 mag) and are faint in the F182M band (F182M $>$ 26.4 mag). As such, they were selected as optically faint dusty galaxy candidates, with high inferred stellar masses. In this study, based on their broad H$\beta$ emission line, ``v-shape" colors, and compact morphology in F444W, we now identify both sources as LRDs, following commonly used criteria (e.g., \citealt{Matthee2024, Labbe2025, Greene2024, Kokorev2024, Kocevski2025}; see details in Sect.~\ref{Sec: lrd}).

\subsection{JWST observations}\label{Sect:fresco} 

The two sources were initially observed by the JWST FRESCO survey \citep{Oesch2023}, which provides NIRCam/grism spectroscopy in the F444W filter and direct imaging in F182M, F210M, and F444W over $\sim$62 arcmin$^2$ in each GOODS field, North and South. The grism spectra cover a wavelength range of 3.8 to 5.0 $\mu$m at a resolution of R$\sim$1,600, reaching a 5$\sigma$ line sensitivity of $\sim1.3\times$10$^{-18}$ \fluxunit. The images reach typical 5$\sigma$ depths of 28.3, 28.1, and 28.2 mag, respectively, in a 0\farcs16 circular aperture. Data reduction was performed using the \texttt{grizli} pipeline\footnote{\url{https://github.com/gbrammer/grizli}}, including continuum subtraction and optimal 1D spectral extraction.

Recently, we obtained deeper NIRCam/grism spectroscopy from a Cycle 3 JWST program (GO-4762; PIs: S. Fujimoto \& G.~Brammer), using the F410M filter instead of F444W. This choice improves spectral sensitivity by blocking the high thermal background beyond 4.4 $\mu$m, at the cost of reduced wavelength coverage. Despite a slightly shorter exposure time (by a factor of 1.3) than FRESCO, this dataset thus achieves a 5$\sigma$ line sensitivity of $\sim7.8\times$10$^{-19}$ \fluxunit. The continuum subtraction was optimized to explicitly mask emission lines and a wide kernel with a large central hole (151, 31 pixels, respectively; for the running median) to prevent over-subtraction of broad wings \citep[see][]{Matthee2024}. 

The NIRCam images and grism spectra of ID9094 and ID2756 are shown in Fig.~\ref{fig1}, updated with the latest data.


\begin{table}
\caption{Flux or flux densities of two LRDs from multi-wavelength observations.}   
\tiny          
\centering
\setlength{\tabcolsep}{1pt}
\renewcommand{\arraystretch}{1.2}
\begin{threeparttable}   
 
\begin{tabular}{l c c cc }    
\hline\hline       
                  
Instrument & $\lambda_{\rm ref}$ & Units & ID2756 & ID9094  \\ 
\hline  
\textit{Chandra}$^{a}$ & 0.5--7 keV & erg\,s$^{-1}$\,cm$^{-2}$ & $<1.1\times 10^{-16}$ & $<1.1\times 10^{-16}$ \\
                 & 2--7 keV  & erg\,s$^{-1}$\,cm$^{-2}$ & $<1.8\times 10^{-16}$ & $<1.8\times 10^{-16}$ \\
                 & 0.5--2 keV  & erg\,s$^{-1}$\,cm$^{-2}$ & $<3.6\times 10^{-17}$ & $<3.6\times 10^{-17}$ \\
\HST/ACS F435W & 0.43\,\textmu m &  nJy & $<19.4$ & $<17.7$  \\
\HST/ACS F606W & 0.59\,\textmu m &  nJy & $<12.3$ & $<9.4$ \\
\HST/ACS F775W & 0.77\,\textmu m &  nJy & $<13.6$ & $<17.7$ \\
\HST/ACS F814W & 0.80\,\textmu m &  nJy & $<17.1$ & $<13.5$ \\
\HST/ACS F850LP & 0.90\,\textmu m & nJy & $<19.5$ & $<24.4$ \\
\HST/WFC3 F105W & 1.06\,\textmu m & nJy & 33.0 $\pm$ 8.4 & ...\\
\HST/WFC3 F125W & 1.25\,\textmu m & nJy & 87.5 $\pm$ 10.8 & 75.7 $\pm$ 9.8 \\
\HST/WFC3 F140W & 1.40\,\textmu m & nJy & 84.7 $\pm$ 6.6 & ...\\
\HST/WFC3 F160W & 1.54\,\textmu m & nJy & 91.3 $\pm$ 7.8 & 109.7 $\pm$ 10.4 \\

\JWST/NIRCam F090W & 0.90\,\textmu m  & nJy & $<14.3$ &$<21.0$ \\
\JWST/NIRCam F115W & 1.15\,\textmu m  & nJy & 79.2 $\pm$ 4.4 & 64.8 $\pm$ 6.2 \\
\JWST/NIRCam F150W &  1.50\,\textmu m & nJy &...& 93.7 $\pm$ 12.9 \\
\JWST/NIRCam F182M &  1.85\,\textmu m & nJy & 80.0 $\pm$ 4.0 & 99.3 $\pm$ 6.7 \\
\JWST/NIRCam F200W & 1.99\,\textmu m  & nJy &...& 109.6 $\pm$ 11.5 \\
\JWST/NIRCam F210M &  2.10\,\textmu m & nJy & 117.8 $\pm$ 5.9 & 130.2 $\pm$ 8.8 \\
\JWST/NIRCam F356W &  3.57\,\textmu m & nJy & 225.8 $\pm$ 11.3 & 443.8 $\pm$ 22.2 \\
\JWST/NIRCam F444W & 4.40\,\textmu m  & nJy & 372.4 $\pm$ 18.6 & 887.7 $\pm$ 44.4 \\

\textit{Spitzer}/IRAC & 3.6\,\textmu m & nJy & 191.0 $\pm$ 15.4  & 400.4  $\pm$ 13.0 \\
                      & 4.5\,\textmu m & nJy & 340.2 $\pm$ 17.6  & 592.7  $\pm$ 16.5 \\
                      & 5.8\,\textmu m & nJy & 933.7 $\pm$ 243.5 & $<676.5$ \\
                      & 8.0\,\textmu m & nJy & 351.2 $\pm$ 284.8 & 2683.0 $\pm$ 289.4   \\

\textit{Spitzer}/MIPS$^{b}$ & 24\,\textmu m &  \textmu Jy & $<21 $ & $<21 $ \\
                      & 70\,\textmu m &  mJy & $<2.4 $ & $<2.4 $ \\
\textit{Herschel}/PACS$^{b}$ & 100\,\textmu m &  mJy & $<1.1$ & $<1.1$ \\
                       & 160\,\textmu m &  mJy & $<2.7$ & $<2.7$ \\
\textit{Herschel}/SPIRE$^{b}$ & 250\,\textmu m & mJy & $<5.7$ & $<5.7$ \\
                        & 350\,\textmu m & mJy & $<7.2$ & $<7.2$ \\
                        & 500\,\textmu m & mJy & $<9$ & $<9$ \\
 SCUBA-2$^{c}$ & 850\,\textmu m & mJy & $<1.2$ & $<1.2$ \\
NOEMA & 1.3\,mm & \textmu Jy & $<151$ & $<91$ \\
      & \cii$_{\rm 158 \mu m}$ & Jy km s$^{-1}$ & $<0.310$ & $<0.157$\\
NIKA-2$^{d}$ & 1.2\,mm & \textmu Jy & $<510$ & $<510$\\
        & 2.0\,mm & \textmu Jy & $<144$ & $<144$\\
VLA 1.5\,GHz$^{e}$ & 20\,cm  & uJy & $<6.6$ & $<6.6$  \\

\hline 
                 
\end{tabular}
\begin{tablenotes}
\item \textbf{Note:} We use 1$\sigma$ for uncertainties and 3$\sigma$ upper limits for non-detections. (a) We calculate 3$\sigma$ upper limits based on the mean sensitivity limits achieved in the central $\sim$1 arcmin$^2$ area at the average aim point from 2 Ms Chandra Deep Field-North survey \citep{xue2016}. (b) Sources are not detected, we adopt 3$\sigma$ upper limits from the GOODS-Herschel survey \citep{Elbaz2011}. (c) We adopt the confusion limit for a 3$\sigma$ detection in the central region of GOODS-N from \cite{cowie2017}. (d) We adopt 3$\sigma$ upper limits according to the N2CLS depth from \cite{Bing2023}. (e) Calculated based on the rms noise for a 1\farcs6 resolution in the field center from \cite{owen2018}.
\end{tablenotes}
\label{table:data} 
\end{threeparttable} 
\end{table}

\begin{table*}
\caption{Best-fit emission line properties from JWST/NIRCam F410M grism spectra, and best-fit UV and optical slopes from photometric bands.} 
\tiny          
\centering
\renewcommand{\arraystretch}{1.2}
\begin{threeparttable} 
 
\begin{tabular}{l c c cccc c c}    
\hline\hline       
ID  & $z_{\rm spec}$ & Flux$_{\rm H\beta, broad}$ & Flux$_{\rm H\beta, narrow}$ & Flux$_{\rm \oi\lambda5008}$  & $\rm FWHM_{\rm H\beta, broad}$ & $\rm FWHM_{\rm narrow}$ & $\beta_{\rm UV}$ & $\beta_{\rm opt}$\\
&&$10^{-18}$erg s$^{-1}$cm$^{-2}$ & $10^{-18}$erg s$^{-1}$cm$^{-2}$ & $10^{-18}$erg s$^{-1}$cm$^{-2}$ & km s$^{-1}$ & km s$^{-1}$\\
\hline  
 ID9094 &  $7.0388\pm0.0001$   & $6.95\pm1.57$ & $2.01\pm0.63$ & $24.47\pm0.49$ &  $2280\pm490$ &  $425\pm10$  & $-1.05 \pm 0.24$ & $1.14 \pm 0.31$ \\
 ID2756 &  $7.1883\pm0.0001$   & $2.65\pm0.65$ & $0.52\pm0.40$ & $6.69\pm0.29$ & $1070\pm240$   &  $295\pm16$  & $-1.48 \pm 0.12$ & $0.26 \pm 0.31$   \\
\hline
\end{tabular}
\label{table2} 
\begin{tablenotes}
\item \textit{Note.} $\beta_{\rm UV}$ and $\beta_{\rm opt}$ are continuum slopes derived from multi-band JWST photometry blueward and redward of the Balmer break, respectively, and trace the two sides of the ``v-shape'' SEDs (see Sect.~\ref{Sec: lrd}).
\end{tablenotes}
\end{threeparttable} 
\end{table*}

\subsection{NOEMA observations}\label{Sect:noema}
The NOEMA observations of ID9094 and ID2756 were carried out in the summer of 2023 using the PolyFiX correlator (Project ID: S23CY, PIs: D. Elbaz \& M. Xiao). Conducted with array configuration D in band 3, the observations targeted the \cii$\,158\,{\rm \mu m}$ emission line, and the 1.3 mm dust continuum, with representative frequencies of 232.052 GHz for ID9094 and 236.576 GHz for ID2756. The total on-source time was 6 hours and 4.5 hours for ID2756 and ID9094, respectively.

All the calibrations and the creation of $uv$ table were performed using CLIC package from the IRAM GILDAS software\footnote{\url{https://www.iram.fr/IRAMFR/GILDAS/}} with the support of IRAM astronomers. We then performed further analysis with version 6.5.5 of the Common Astronomy Software Application package \citep[CASA;][]{McMullin2007}. 

For the \cii$\,158\,{\rm \mu m}$ emission line, imaging was carried out using the \texttt{tclean} task with 0.4$^{\prime\prime}$ pixels and a channel width of 50 km s$^{-1}$ with a natural weighting. The natural weighting provides the best point-source sensitivity, which is optimal for source detection.
The resulting data cube has a synthesized beam size of full width at half maximum (FWHM) $\simeq$ 1\farcs838 $\times$ 1\farcs472 ($\sim$9.8 kpc $\times$ 7.8 kpc in physical scale) and 2\farcs859 $\times$ 1\farcs854 ($\sim$ 15.0 kpc $\times$ 9.7 kpc) with a root mean square (rms) sensitivity of $\sim$ 0.45 and 0.84 mJy beam$^{-1}$ per channel at the phase center, for ID9094 and ID2756, respectively. 

We also created the observed 1.3 mm continuum maps using natural weighting and the same method as for the \cii$\,158\,{\rm \mu m}$ emission line. For ID9094, the rms level is $\sim$ 30.4 $\mu$Jy beam$^{-1}$ in the map of 1\farcs725 $\times$ 1\farcs383 ($\sim$ 9.2 kpc $\times$ 7.4 kpc) angular resolution. For ID2756, the rms level is $\sim$ 50.3 $\mu$Jy beam$^{-1}$ in the map of 2\farcs712 $\times$ 1\farcs743 ($\sim$ 14.3 kpc $\times$ 9.2 kpc) angular resolution.

\subsection{Multi-wavelength observations}
Besides the JWST FRESCO and NOEMA observations, the two sources are also covered by a wide array of multi-wavelength observations. In Table \ref{table:data}, we summarize the multi-wavelength dataset for our two sources from X-ray to radio: ($i$) X-ray: \textit{Chandra} 2 Ms (0.5–7.0keV, 0.5–2.0keV, and 2–7keV bands) images in the Chandra Deep Field-North field \citep[CDF-N;][]{xue2016}; ($ii$) \HST/ACS (F435W, F606W, F775W, F814W, F850LP) and HST/WFC3 (F105W, F125W, F140W, F160W) images from the \textit{Hubble} Legacy Fields Program \citep[HLF\footnote{\url{https://archive.stsci.edu/prepds/hlf/}};][]{Whitaker2019}; ($iii$) \JWST/NIRCam (F090W, F115W, F150W, F182M, F200W, F210M, F356W, F444W) from FRESCO  \citep{Oesch2023}, CONGRESS \citep{Egami2023}, and JADES \citep{Eisenstein2023, D_Eugenio2024}. The two sources are not yet covered by any publicly available \JWST/MIRI observations; ($iv$) \textit{Spitzer}/IRAC from the GREATS program \citep[3.6 \,\textmu m, 4.5 \,\textmu m, 5.8 \,\textmu m, 8 \,\textmu m;][new catalogs based on FRESCO sources are taken from Stefanon et al. in prep; priv. comm.]{Stefanon2021}; ($v$) \textit{Spitzer}/MIPS, \textit{Herschel}/PACS and SPIRE catalog is taken from the GOODS-Herschel survey \citep{Elbaz2011}; ($vi$) JCMT/SCUBA-2 at 850 \,\textmu m \citep{cowie2017}; ($vii$) NOEMA data from this study (see Sect. \ref{Sect:noema}); ($viii$) NIKA2 Cosmological Legacy Survey at 1.2 mm and 2.0 mm \citep{Bing2023}; and ($ix$) VLA 1.5 GHz \citep{owen2018}.

\begin{figure*}
\centering
\includegraphics[width=18cm]{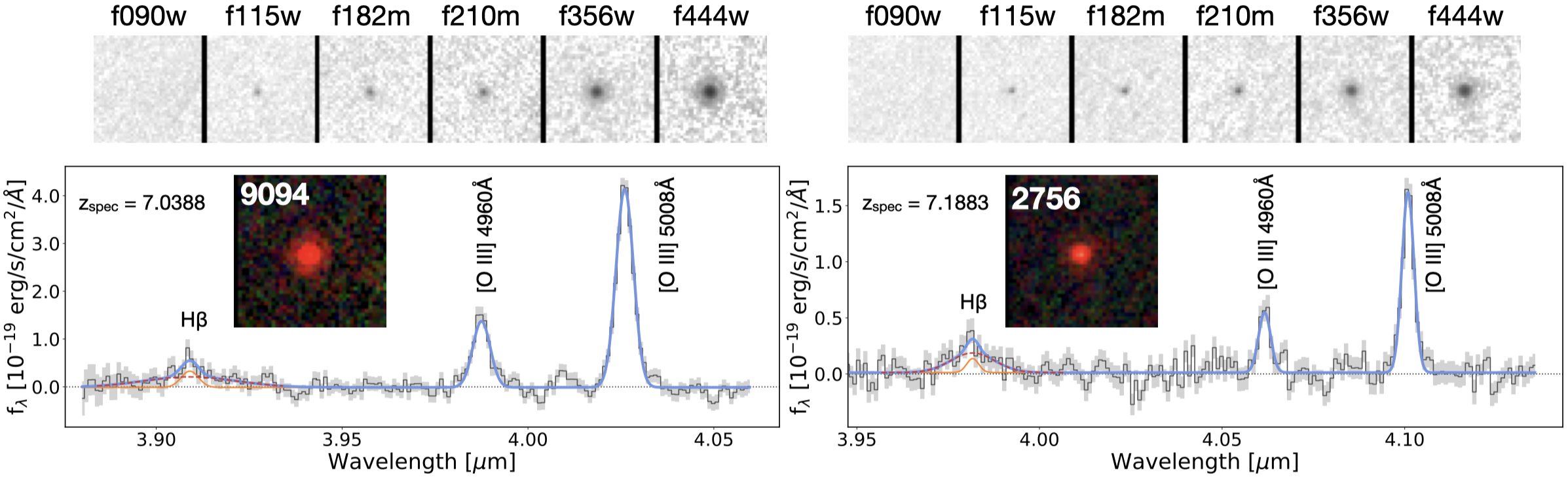}
\caption{\textbf{JWST images and spectra of the two sources}. \textit{Top}: 2$^{\prime\prime}$ $\times$ 2$^{\prime\prime}$ stamps obtained in JWST/NIRCam filters (0.90$\mu$m,1.15$\mu$m,1.82$\mu$m, 2.10$\mu$m, 3.56$\mu$m, and 4.44$\mu$m).
\textit{Bottom}: 1D spectra (covering H$\beta$, \oi\,$\lambda\lambda4960,5008\AA$ emission lines) obtained from NIRCam/grism observations with the F410M filter, with RGB images embedded (F182M in blue, F210M in green, and F444W in red). The gray shaded areas show the associated 1$\sigma$ uncertainty. The best-fit Gaussian line model is shown in blue. Both sources have a broad H$\beta$ line, where the solid orange line shows the narrow component and the dashed red line shows the broad component. 
      }
         \label{fig1}
\end{figure*}

\begin{figure*}
\centering
\includegraphics[width=17cm]{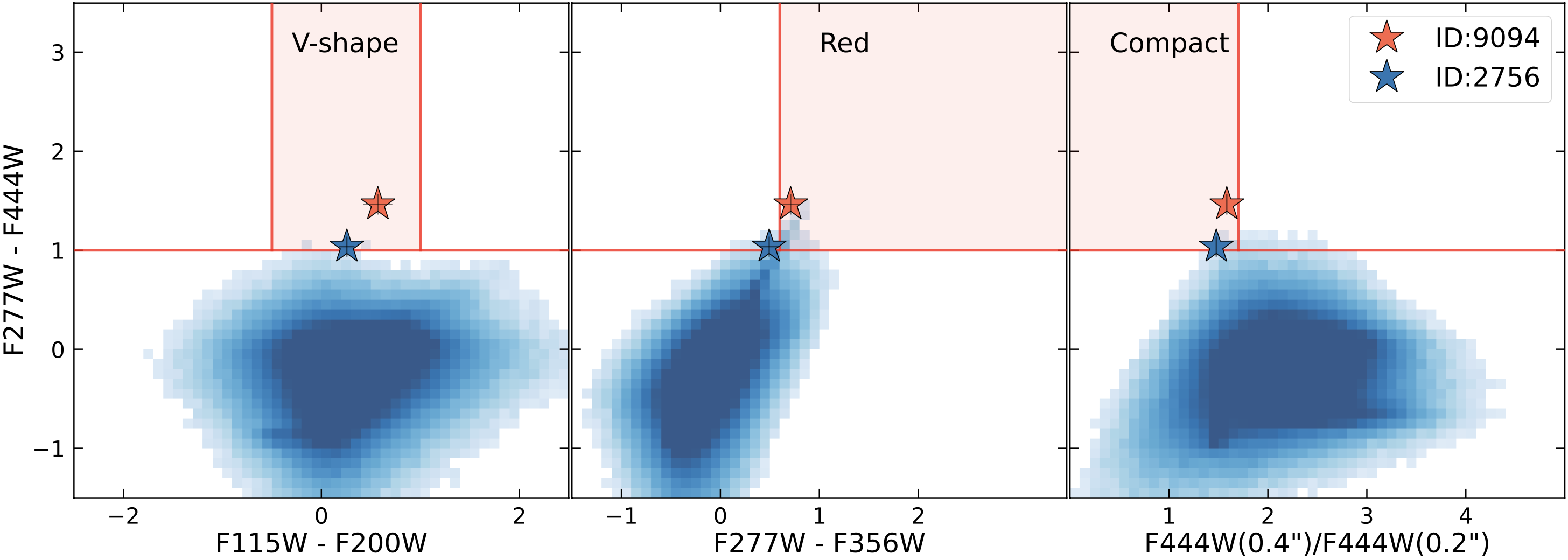}
\caption{\textbf{The locations of the two sources (red and blue stars) relative to the typical color and compactness selection criteria for LRDs.} The red-shaded regions highlight LRDs exhibiting ``v-shape" SED (\textit{left}), red (\textit{middle}), and compact morphology (\textit{right}), defined by the criteria: $-0.5< \rm{F115W}-\rm{F200W} < 1.0$, $\rm{F277W}-\rm{F444W} > 1.0$, $\rm{F277W}-\rm{F356W} > 0.6$, and $f_{\rm F444W}(0\farcs4)/f_{\rm F444W}(0\farcs2) < 1.7$ (see Sect. \ref{Sec: lrd}). Error bars indicate 1$\sigma$ uncertainties. The blue-shaded region represents the distribution of sources observed in these filters from the prime JWST blank legacy fields in the DJA.
      }
         \label{fig1_}
\end{figure*}

We note that the photometric measurements from HST and JWST images are derived following the same procedure as outlined in \citet{Weibel2024}. Briefly, we use \texttt{SExtractor} \citep{Bertin1996} in dual image mode with an inverse-variance weighted stack of the F210M+F444W with FRESCO-only imaging, as the detection image. In this study, fluxes are measured in 0\farcs16 radius circular apertures in images that are point spread function (PSF)-matched to the F444W band. Total fluxes are derived from the Kron AUTO aperture provided by \texttt{SExtractor} in the F444W band, in addition to a correction based on the encircled energy of the Kron aperture on the F444W PSF. Detailed descriptions of data reduction and photometric measurements are provided in \cite{Weibel2024} and \cite{Xiao2024}.

\begin{figure*}
\centering
\includegraphics[width=17cm]{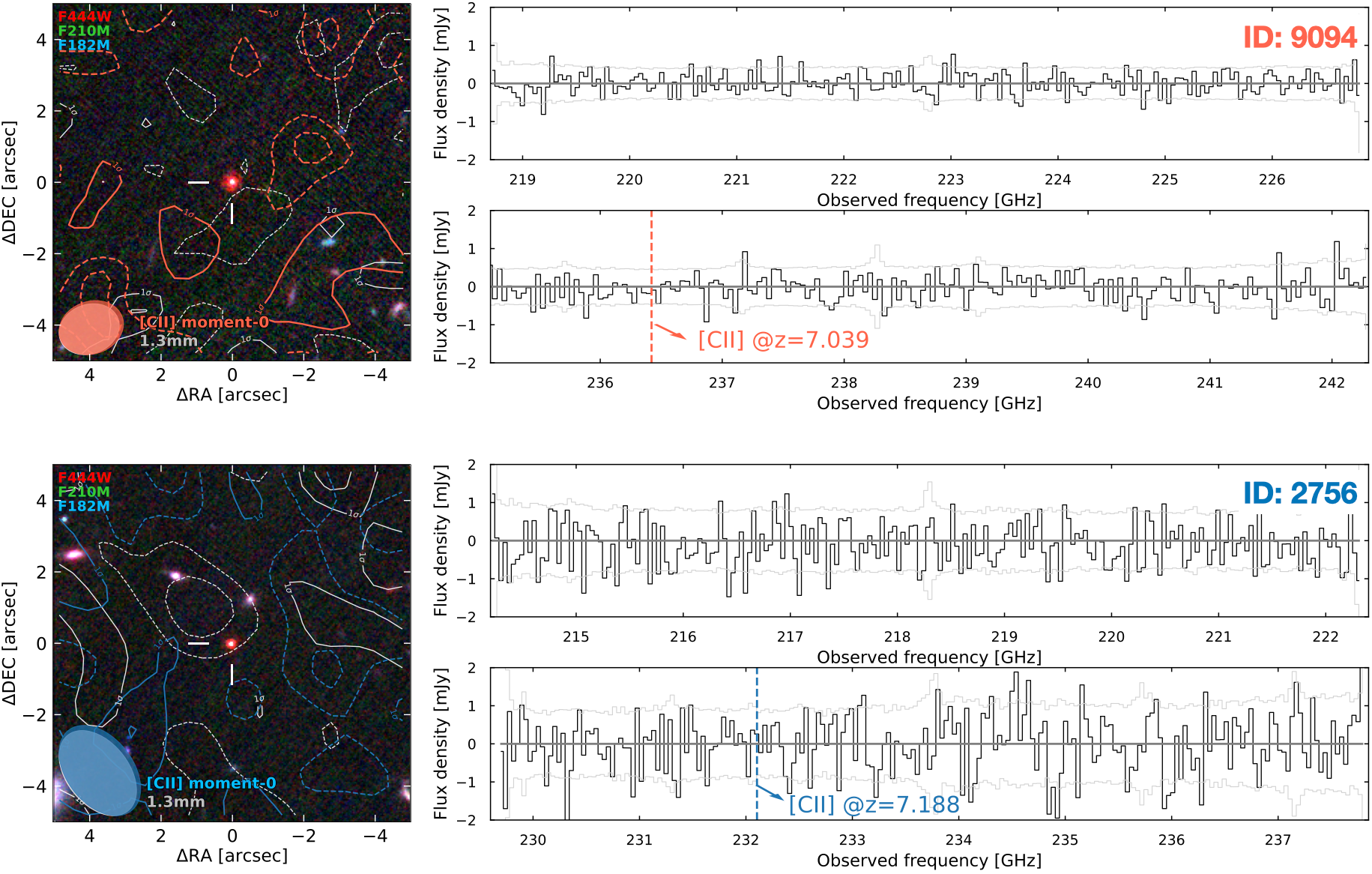}
\caption{\textbf{No detection of the \cii$\,158\,{\rm \mu m}$ emission line and 1.3 mm dust continuum with NOEMA observations}. \textit{Left}: \cii$\,158\,{\rm \mu m}$ line moment-0 map (assuming line width of 250 km s$^{-1}$) and 1.3 mm dust contours overlaid on the JWST RGB image ($10^{\prime\prime}\times10^{\prime\prime}$). The contour levels start at 1$\sigma$ and increase in steps of $\pm1\sigma$, where positive and negative contours are solid and dashed, respectively. The beam sizes are displayed in the lower left corner. \textit{Right}: NOEMA 1.3 mm spectra with 50 km s$^{-1}$ binning. The grey lines show the associated 1$\sigma$ uncertainty. The vertical dashed lines highlight the locations of \cii$\,158\,{\rm \mu m}$. The spectra are taken at the positions of the sources and match the apertures of the beam sizes, assuming the emission line is unresolved. 
      }
         \label{noema_nodetection}
\end{figure*}

\subsection{Our sample: two ``Little Red Dots” at $z_{\rm spec}>7$}\label{Sec: lrd}
Although there is no uniform definition of LRDs, we identify our two sources as LRDs because they satisfy both of two commonly used selection criteria: 1) an initial criterion -- sources with broad Balmer emission lines \citep[$\nu_{\rm FWHM, broad} >1000$ km s$^{-1}$;][]{Matthee2024}; and 2) a widely used criterion based only on photometric measurements -- a combination of ``v-shape" SED and compactness selection \citep[e.g.,][]{Labbe2025, Kokorev2024, Greene2024, Kocevski2025}.

Fig.~\ref{fig1} presents the JWST/NIRCam images and grism spectra of ID9094 and ID2756. The RGB images show that both sources appear red with point-like morphology. The grism spectra show H$\beta$ and \oi$\lambda\lambda4960,5008$ emission lines. Both sources exhibit a broad H$\beta$ line. We fit the H$\beta$ line with both narrow and broad Gaussian components, assuming the narrow component has the same width as the \oi$\lambda\lambda4960,5008$ lines. The best-fit line properties are shown in Table~\ref{table2}. The broad component of H$\beta$ has a FWHM of $\nu_{\rm FWHM, H\beta, broad} = 2280\pm490$ km s$^{-1}$ for ID9094 and $\nu_{\rm FWHM, H\beta, broad} = 1070\pm240$ km s$^{-1}$ for ID2756, suggesting a possible Type I (broad emission line) AGN. Notably, ID2756 has also been identified as a strong Lyman-$\alpha$ emitter, with a considerable equivalent width of 221 \AA~(Leonova et al. in prep.). 

\begin{figure*}
\centering
\includegraphics[width=17cm]{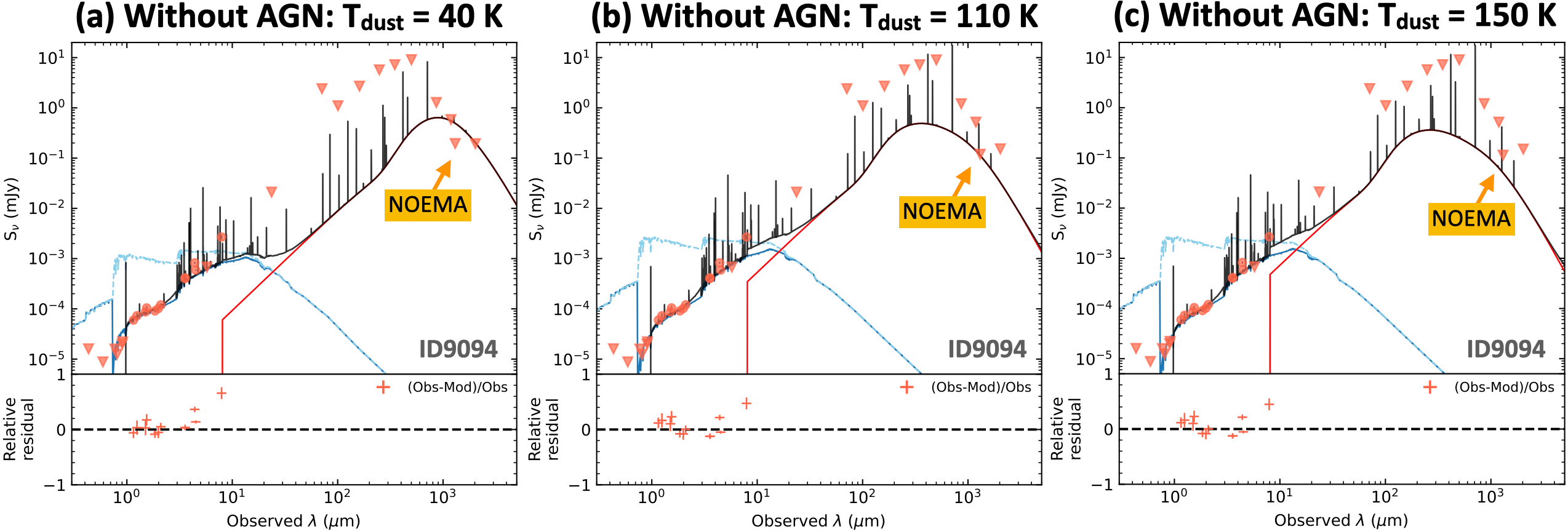}
\includegraphics[width=17cm]{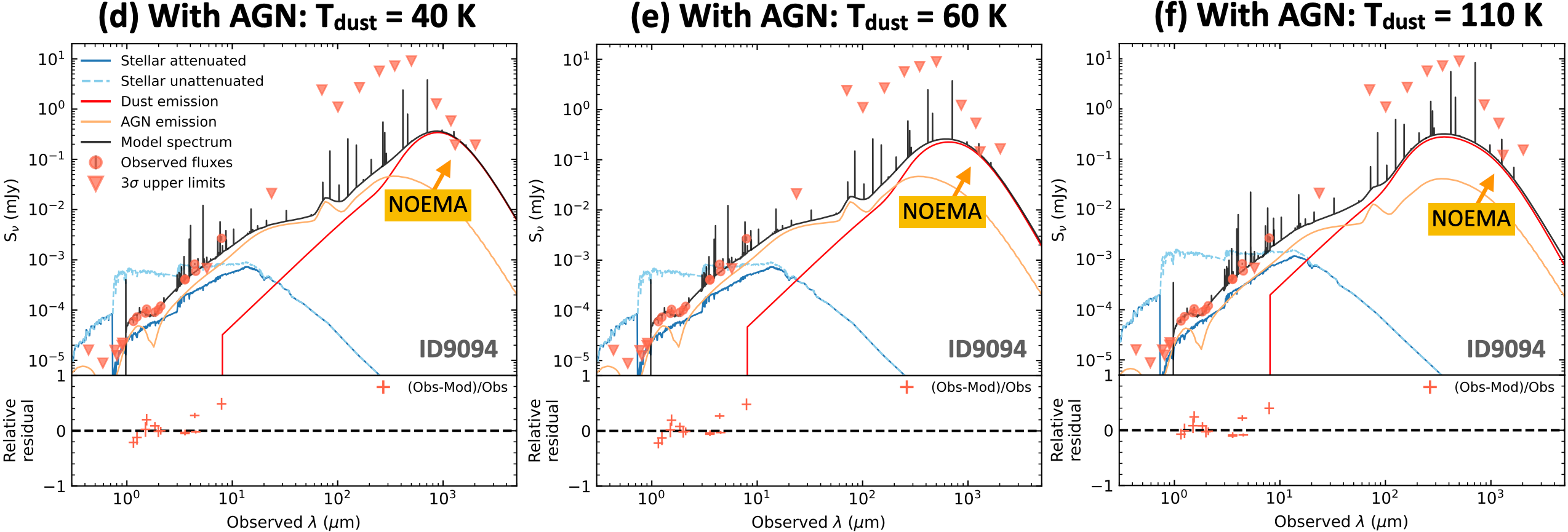}
\caption{\textbf{Best-fit SEDs of ID9094 with fixed dust temperatures, providing the first hint of an AGN scenario in this work}. \textit{Top panels}: Best-fit SEDs without an AGN component; \textit{bottom panels}: Best-fit SEDs with an AGN component. The 3$\sigma$ upper limit from NOEMA 1.3 mm, the deepest FIR photometric constraint, is highlighted, playing a key role in shaping the FIR SED and constraining the lower limit of $T_{\rm dust}$. The FIR photometry has been corrected for CMB effects. We show three representative $T_{\rm dust}$ during the fitting with and without AGN component, which results in the predicted 1.3 mm flux exceeding ($a$ and $d$), same as ($b$ and $e$), below ($c$ and $f$) the observed 3$\sigma$ upper limit. Panel $d$ shows the fit with AGN and $T_{\rm dust} = 40$ K, which slightly exceeds the 3$\sigma$ limit but is not ruled out given the model degeneracy and observational uncertainties (see Sect.~\ref{Sec: sed}). Without an AGN, the SED fitting requires $T_{\rm dust} \,\gtrsim\, 110$ K. This value significantly exceeds the typical $T_{\rm dust}\sim40$ K observed in REBELS galaxies at similar redshifts \citep{Sommovigo2022}, and may indicate additional dust heating mechanisms such as AGN activity. In contrast, including an AGN component (\textit{bottom panels}) allows acceptable fits with $T_{\rm dust}$ as low as $\sim$40-60 K, similar to AGN and/or quasar host galaxies at $z>4$ \citep[$\sim40-100$ K; e.g.,][]{Walter2022,Tsukui2023,Tripodi2023, Decarli2023}, supporting the presence of an AGN in ID9094. }
         \label{9094_sed}
\end{figure*}

We further compare our two sources to an alternative color and compactness selection criterion for LRDs, designed to identify compact, red objects with ``v-shape" SEDs \citep[e.g.,][]{Kokorev2024,Greene2024}:
\begin{eqnarray*}
    && \text{S/N(F444W)} > 14\quad \&\quad \text{m}_{\text{F444W}} < 27.7 \text{ mag},\\
    && -0.5 < \text{F115W} - \text{F200W} < 1.0 , \\
    && \text{F277W} - \text{F444W} > 1.0 , \\
    && \text{F277W} - \text{F356W} > 0.7, \\
    && \text{compact}=f_{\rm F444W}(0\farcs4)/f_{\rm F444W}(0\farcs2) < 1.7.
\end{eqnarray*}
Both sources are bright in F444W, with magnitudes of $24.0\pm0.1$ mag (ID9094) and $25.0\pm0.1$ mag (ID2756), consistent with the LRD selection. Since ID2756 lacks observations in F277W and F200W, while ID9094 is missing F277W, the corresponding fluxes used here are derived from the best-fit SED and the 16th–84th percentile range (as shown later in Fig.~\ref{9094_sed}). The detailed comparison is shown in Fig.~\ref{fig1_}. ID9094 lies within the red region, perfectly matching the LRD selection. ID2756 exhibits a "v-shaped" color and compact morphology, aligning with the LRD classification. Its F277W – F356W color is slightly bluer than the selection threshold but still within the 1$\sigma$ uncertainty. We therefore classify both sources as LRDs. Compared to the majority of sources from various JWST blank legacy fields (blue-shaded region in Fig.~\ref{fig1_}), obtained from the DAWN JWST Archive \citep[DJA\footnote{\url{https://dawn-cph.github.io/dja/imaging/v7/}};][]{Valentino2023}, these two LRDs stand out as particularly exceptional, exhibiting redder colors and more compact morphologies.

To further quantify the “v-shape” SEDs, we also measured the UV and optical continuum slopes ($\beta_{\rm UV}$ and $\beta_{\rm opt}$) using multi-band photometry, following the method of \cite{Kocevski2025}. Both sources satisfy their LRD selection criteria ($-2.8<\beta_{\rm UV} < -0.37$ and $\beta_{\rm opt} > 0$). The measured slope values are listed in Table~\ref{table2}. In addition, both sources are unresolved in F444W, further supporting that they are typical LRDs.

Altogether, we find that both sources are consistent with standard criteria for LRDs based on their morphology and ``v-shape" SED. In addition, both sources show the broad H$\beta$ line, making them strong Type I AGN candidates.

\section{Results}\label{Sec: result}
\subsection{Non-detections in both \cii$\,158\,{\rm \mu m}$ and 1.3 mm of two LRDs} \label{Sec: no_detection}

We analyze the NOEMA data cubes and extract spectra at the positions of our sources using an aperture consistent with the beam size. The resulting spectra, shown in Fig.~\ref{noema_nodetection}, reveal no significant signal at the expected \cii\ frequency ($\ll3\sigma$). Assuming a \cii$\,158\,{\rm \mu m}$ line width of 250 km s$^{-1}$ -- a typical value for high-redshift galaxies \citep{bethermin2020, Endsley2022} -- we construct the velocity-integrated intensity map (moment-0). Fig.~\ref{noema_nodetection} overlays the contours from the \cii\ moment-0 and 1.3 mm maps on the JWST RGB image, showing no detections ($\ll3\sigma$) at the source positions and in the surrounding regions. We therefore conclude that neither the \cii$\,158\,{\rm \mu m}$ emission line nor the 1.3 mm dust continuum emission are detected for the two LRDs in the current NOEMA data.

We then determine 3$\sigma$ upper limits for both \cii\ and dust continuum. Using the rms of the 1.3 mm map (see Sect.~\ref{Sect:noema}) and assuming our sources are not resolved, we derive 3$\sigma$ upper limits of 91 $\mu$Jy and 151 $\mu$Jy for ID9094 and ID2756, respectively. 
In addition, the 3$\sigma$ upper limits for \cii\ are calculated as three times the pixel-by-pixel rms of the moment-0 map, which gives upper limits of 0.157 Jy km s$^{-1}$ and 0.310 Jy km s$^{-1}$ for ID9094 and ID2756, respectively. The corresponding 3$\sigma$ upper limits on the \cii\, emission-line luminosity are log($L_{\rm \cii}/L_\odot) < 8.23$ for ID9094 and log($L_{\rm \cii}/L_\odot) < 8.58$ for ID2756. Using $L_{\rm \cii}$, we further calculate the 3$\sigma$ upper limits on the molecular gas mass \citep{zanella2018} to be log($M_{\rm mol}/M_\odot)<9.7$ for ID9094 and log($M_{\rm mol}/M_\odot)<10.1$ for ID2756.

Intriguingly, when requesting the NOEMA observations, we assumed these two galaxies to be DSFGs. Given their stellar masses from \cite{Xiao2024}, we estimated their total star formation rates (SFR$_{\rm tot}$) assuming that they lie on the star-forming main sequence \citep[SFMS;][]{Schreiber2015} and predicted their expected \cii\ fluxes using the $L_{\rm \cii}$–SFR$_{\rm tot}$ relation for SFGs \citep{schaerer2020}. Similarly, we predicted their 1.3 mm continuum fluxes by assuming an infrared SED shape consistent with typical dust-obscured galaxies \citep[optically dark/faint galaxies;][]{Xiao2023}. Given the depth of our NOEMA observations, we expected ID9094 to be detected at $>16\sigma$ in \cii\ and $>3\sigma$ in the 1.3 mm continuum, while ID2756 should be detected in \cii\ at $>5\sigma$. However, the complete absence of \cii\ and 1.3 mm emission suggests that these two LRDs are not the typical dusty SFGs initially assumed. This result aligns with our findings in Sect.~\ref{Sec: lrd}, where both sources exhibit a broad H$\beta$ emission line, being Type-I AGN candidates.

\subsection{The hints of AGNs in two LRDs from FIR non-detections }
In this section, we utilize the 3$\sigma$ upper limits on the \cii$\,158\,{\rm \mu m}$ emission line and the 1.3 mm continuum to investigate the possible presence of AGN activity in the LRDs. While both sources have already been identified as strong Type-I AGN candidates based on their broad H$\beta$ emission lines, our primary goal here is to investigate whether the FIR non-detections can effectively rule out the DSFG scenario for LRDs, independently of spectral information. Here, we focus primarily on ID9094, which benefits from deeper NOEMA observations compared to ID2756, allowing for more robust constraints. We assess the AGN contribution through two independent approaches: 1) UV-to-FIR spectral energy distribution (SED) fitting analysis and 2) the $L_{\rm \cii}$–SFR$_{\rm tot}$ empirical relation.

\subsubsection{SED fitting analysis}\label{Sec: sed}
We perform a UV-to-FIR SED fitting analysis using \texttt{CIGALE} \citep{Boquien2019}, which is based on an energy balance principle, fixing the redshift to $z_{\rm spec}$. We assume a constant star formation history (SFH) and adopt \citet{Bruzual2003} stellar population models with the \citet{Calzetti2000} dust attenuation law. We adopt a broad metallicity range of 0.004 to 0.02, dust attenuation to the rest-frame V band ($A_{\rm V}$) values from 0 to 6 magnitudes, and ionization parameters log$U$ of -2.0. The AGN component is modeled using the \citet{Fritz2006} template, with the AGN fraction ($f_{\rm AGN}$), i.e., the contribution of the AGN to the total IR luminosity, ranging from 0.01 to 0.8. The angle between the equatorial axis and the viewing angle at the line of sight ($\rm \Psi$) ranges from 40 to 90. With a fixed open angle of the AGN torus at 100 degrees, this range of $\rm \Psi$ refers to Type I AGN. We note that recent MIR observations \citep{Wangbingjie2025,Williams2024_RED,Barro2024} suggest a lack of hot torus dust emission in LRDs, which may deviate from the \citet{Fritz2006} template adopted. However, with the absence of deep MIR observations, we anticipate little change in the quality of SED fitting on our sample by only changing the AGN torus emission in the templates. On the contrary, the inclusion of continuum from accretion disk by limiting $\Psi$ to Type I AGN is critical, as we robustly detected broad lines in H$\rm \beta$. 

Given that we only have upper limits in the FIR, we adopt a simple FIR dust model from \cite{Casey2012}, using the default dust emissivity index ($\beta=2.0$) and MIR power-law slope ($\alpha = 2.0$). Compared to $\beta=1.5$, which is adopted in some studies of dusty star-forming galaxies \citep[e.g.,][]{Hildebrand1983, Kovacs2006, Gordon2010, Gomez-Guijarro2022a, Xiao2023}, our choice of $\beta=2.0$ is more conservative and leads to slightly lower inferred dust temperatures for a given FIR slope. During the fitting process, we find that when the dust temperature ($T_{\rm dust}$) is allowed to vary across a defined range, the best-fit $T_{\rm dust}$ is strongly influenced by the upper boundary of the input range. This effect arises due to the absence of deep MIR data, which prevents us from constraining the maximum $T_{\rm dust}$. Consequently, we cannot determine a precise best-fit value for $T_{\rm dust}$, but the deep NOEMA data allow us to place a lower limit.

To refine our constraints, we adopt an optimized fitting approach that allows us to constrain the minimum allowed $T_{\rm dust}$ using the NOEMA 1.3 mm 3$\sigma$ upper limits. In this approach, we fix $T_{\rm dust}$ during the fitting and perform multiple fits with different $T_{\rm dust}$ values, ranging from 30 K to 150 K. In addition, we also correct the effect of the cosmic microwave background (CMB) as an observing background on photometry, following \cite{dacunha2013}. The CMB has the effect of raising the apparent dust temperature with increasing redshift, as the CMB temperature scales with redshift and provides an additional background source of radiation for dust heating. We assume $\beta=2.0$, consistent with previous studies for $z\sim7$ galaxies \citep{Sommovigo2022}. In this way, we derive the intrinsic FIR photometry starting from MIPS 70$\mu$m, which is used for SED fitting. The best-fit SEDs, with and without AGN components considered, are presented in Fig.~\ref{9094_sed}.

For ID9094, without including an AGN component (Fig.\ref{9094_sed}-$top$), we present three representative examples of best-fit SEDs with ($a$) $T_{\rm dust}=40$ K, ($b$) $T_{\rm dust}=110$ K, and ($c$) $T_{\rm dust}=150$ K. These correspond to results that the predicted flux density at 1.3 mm exceeds, matches, and falls below the observed 3$\sigma$ upper limits, respectively, under the energy balance assumption of \texttt{CIGALE}. At $T_{\rm dust}=40$ K -- a typical value for star-forming UV-bright galaxies in the REBELS survey at $z\sim7$ \citep{Sommovigo2022} -- the predicted 1.3 mm flux exceeds the observed 3$\sigma$ upper limit by nearly a factor of three. This discrepancy suggests that, with our current NOEMA observations, we should have detected the source at a significance level of $\sim8\sigma$, which is clearly not the case. As $T_{\rm dust}$ increases, the peak of the FIR SED moves towards the shorter wavelength and the fit improves. At $T_{\rm dust}=110$ K (Fig.\ref{9094_sed}-$b$), the predicted 1.3 mm flux aligns with the observed 3$\sigma$ upper limit. However, given that the actual signal-to-noise ratio at 1.3 mm is well below 3$\sigma$, this implies that the real $T_{\rm dust}$ would be greater than 110 K.

Therefore, without including an AGN component, we obtain a lower limit of $T_{\rm dust} \,\gtrsim\, 110$ K. At fixed $T_{\rm dust} = 110$ K, as in Table~\ref{table3}, the best-fit IR luminosity is $L_{\rm IR}= 13.8^{+1.8}_{-2.6}\times10^{11} L_\odot$, $\rm{SFR}=125.2^{+8.7}_{-8.6} ~M_{\odot}$yr$^{-1}$ (we use this for the analysis in Sect.~\ref{Sec: relation}), and stellar mass of $M_{\star}=2.3^{+0.2}_{-0.2}\times 10^{10} ~M_{\odot}$. Remarkably, even this lower limit is significantly higher than the typical $T_{\rm dust} \sim 40$ K observed in REBELS galaxies at similar redshifts \citep{Sommovigo2022}. Similar findings have been reported in studies of 675 LRDs at $z\gtrsim4$, regardless of heating mechanism, whether AGN or star formation, where an average luminosity-weighted dust temperature of $T_{\rm dust}=110$ K is reported \citep{Casey2024}. Such an extreme dust temperature is challenging to explain solely through heating by star formation. In typical star-forming galaxies, the dust temperature is primarily driven by the ultraviolet radiation field of young stars, which is usually insufficient to reach the high $T_{\rm dust}$ values observed here \citep[see e.g.,][]{Harshan2024, laporte2017a}.

Based on the best-fit $L_{\rm IR}$ without AGN at fixed $T_{\rm dust} = 110$ K, we derived the dust emitting region size $R_{\rm d}\sim75$ pc, based on a black-body assumption\footnote{$L_{\rm IR} \sim 4\pi R_{\rm d}^2 \sigma T_{\rm dust}^4$ and $\sigma$ is the Stefan–Boltzmann constant.}. A crucial parameter to consider is the stellar surface density, given by $\Sigma_{\star} = M_{\star}/2\pi R_{\rm d}^2 \sim 10^6 M_{\odot}$ pc$^{-2}$ \citep[similar findings are reported in, e.g.,][]{Baggen2024}. This is about one order of magnitude higher than the densest systems in the local universe \citep{Hopkins2010} and SFGs at $z\sim5-14$ observed by JWST \citep{Morishita2024,Schaerer2024}, highlighting a general issue for LRD models that rely solely on stellar components. Expanding the star formation region would reduce $\Sigma_{\star}$, but it would also decrease $T_{\rm dust}$ below the observed values \citep[e.g.,][]{deRossi2018}, leading to an inconsistency with our findings. These values are difficult to explain through star formation alone under standard assumptions and may suggest the contribution of additional heating sources, such as AGN activity. However, we emphasize that this interpretation remains model-dependent, and we further discuss such caveats in Sect.~\ref{Sec:caveats}.


\begin{table*}
\caption{\texttt{CIGALE} best-fit results for ID9094: with and without AGN component.}   
\tiny          
\centering
\renewcommand{\arraystretch}{1.2} 
\begin{threeparttable} 
 
\begin{tabular}{l c c ccc  }    
\hline\hline       
                    
ID9094  & fixed $T_{\rm dust}$ [K] & log $M_{\rm \star} [M_{\odot}$] & \rm{SFR} [$M_{\odot}$yr$^{-1}$]  & $L_{\rm IR}$ [$L_\odot$]\\ 
\hline  
 Without AGN    & 110     &       $2.3^{+0.2}_{-0.2}\times 10^{10}$ &  $125.2^{+8.7}_{-8.6}$ &   $13.8^{+1.8}_{-2.6}\times10^{11}$     \\
 With AGN    &   60  &       $9.7^{+3.0}_{-8.6}\times 10^9$ &  $33.1^{+11.4}_{-12.9}$ &   $6.2^{+2.2}_{-4.3}\times10^{11}$  \\
   With AGN    &   40  &   $4.1^{+5.8}_{-2.9}\times 10^9$ &  $22.4^{+10.8}_{-3.9}$ &  $5.0^{+3.2}_{-1.6}\times10^{11}$  \\
\hline
\end{tabular}
\label{table3} 
\end{threeparttable} 
\end{table*}

When the AGN component is included in the SED fitting (Fig.~\ref{9094_sed}-$bottom$), the results become more physically plausible. Similar to Fig.\ref{9094_sed}-$top$, we also show three representative examples of best-fit SEDs with ($d$) $T_{\rm dust}=40$ K, ($e$) $T_{\rm dust}=60$ K, and ($f$) $T_{\rm dust}=110$ K. However, due to the degeneracy between the dust temperature and the AGN fraction, the current upper limit at 1.3 mm does not allow us to place a meaningful lower limit on $T_{\rm dust}$. The observed photometry is consistent with SED fits with $T_{\rm dust}=60$ K and even marginally consistent with $T_{\rm dust}$ down to $\sim$40 K, a value comparable to those observed in AGN and/or quasar host galaxies at $z>4$ \citep[$\sim40-100$ K; e.g.,][]{Walter2022,Tsukui2023,Tripodi2023, Decarli2023, Meyer2025}. At fixed $T_{\rm dust} = 60\, (40)$ K, the best-fit AGN fraction is $f_{\rm AGN} = 0.44\pm0.12\, (0.58\pm0.12)$, indicating a scenario where both AGN and star formation contribute significantly to the FIR emission. The corresponding best-fit $L_{\rm IR}$, $\rm{SFR}$, and $M_{\star}$ are listed in Table~\ref{table3}. We also note that both sources are covered by the MEOW survey (JWST Cycle 3 GO-5407; PI: G. Leung; priv. comm.), which detects both in F1000W and ID9094 also in F2100W. Although the fluxes are not yet publicly released, preliminary analysis suggests that ID9094 exhibits significant hot dust emission in the MIR (G. Leung, priv. comm.).

For ID2756, we perform SED fitting using the same strategy as for ID9094 (see Appendix~\ref{appendix1}). However, the current 1.3 mm observations for ID2756 (rms level $\sim$50.3 $\mu$Jy beam$^{-1}$) are not sufficiently deep to constrain the SED effectively. The lack of sensitivity prevents us from distinguishing between the DSFG and AGN scenarios based solely on the SED fitting results. To achieve meaningful constraints, deeper FIR/submillimeter observations are required. Specifically, observations at least as deep as for ID9094 (rms level $\sim$30.4 $\mu$Jy beam$^{-1}$) are necessary to provide tighter limits on the 1.3 mm flux density and help refine the SED fitting.

\subsubsection{$L_{\rm \cii}$–SFR$_{\rm tot}$ relation}\label{Sec: relation}
In this section, we compare the locations of our LRDs with the well-established $L_{\rm \cii}$–SFR$_{\rm tot}$ scaling relation to investigate their alignment with expectations for star-forming galaxies or the potential influence of AGN activity. The $L_{\rm \cii}$–SFR$_{\rm tot}$ relation has been shown to hold across a wide range of redshifts, with no significant evolution observed up to $z\sim9$ \citep[e.g.,][]{Lagache2018, schaerer2020, carniani2020, Pallottini2022}. Additionally, studies indicate that the relation is consistent for both AGN-hosting galaxies and typical star-forming galaxies, suggesting that the presence of AGN does not significantly alter the global $L_{\rm \cii}$–SFR$_{\rm tot}$ relation \citep[e.g.,][]{Herrera-Camus2018}.

\begin{figure}
\centering
\includegraphics[width=9cm]{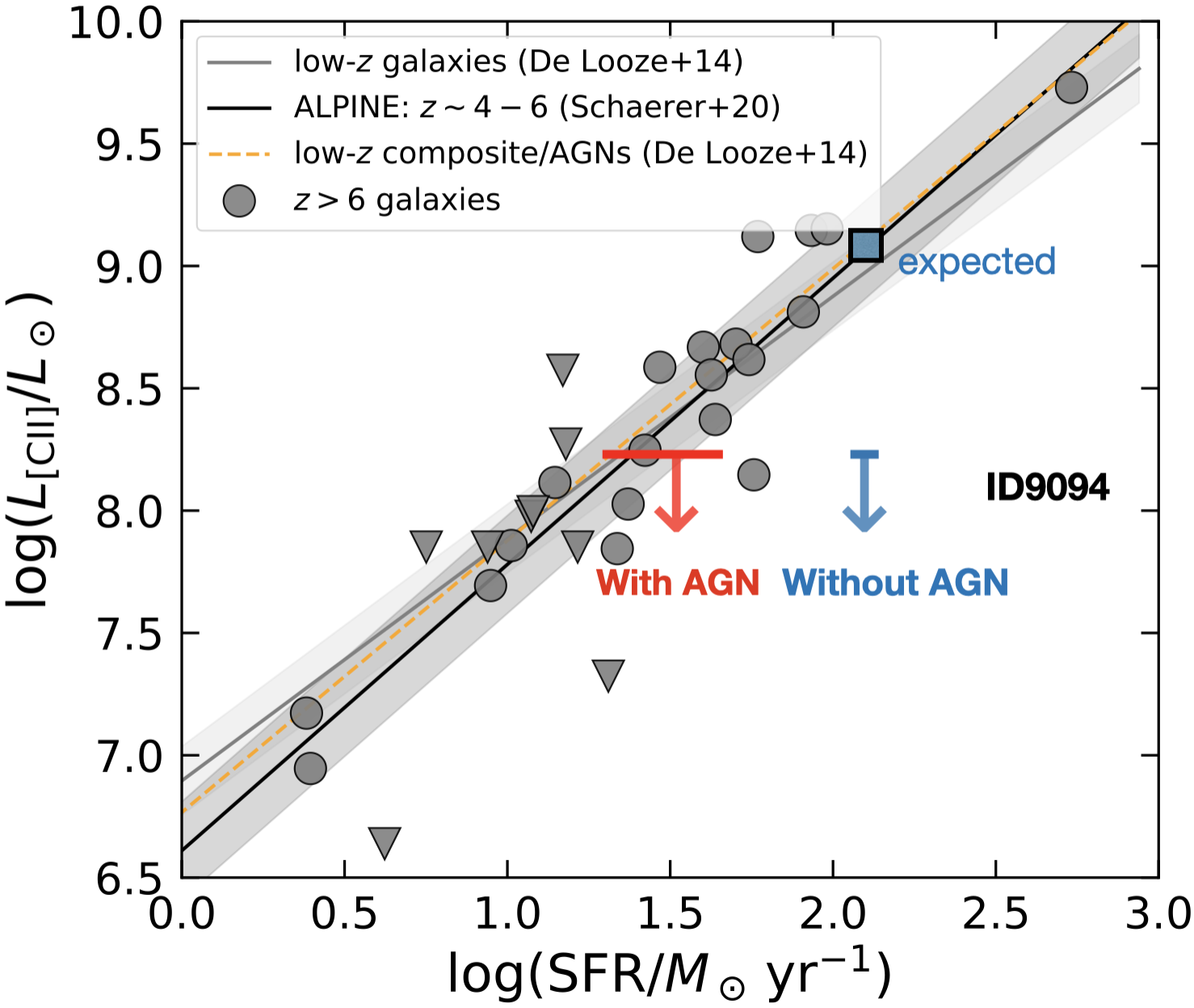}
\caption{\textbf{[CII] luminosity as a function of SFR, providing the second hint of an AGN scenario in this work.} ID9094 is shown with 3$\sigma$ upper limits on $L_{\rm \cii}$ ($L_{\rm \cii} < 1.7 \times 10^{8} L_\odot$) in red and blue arrows, representing cases with (fixed $T_{\rm dust} = 60$ K) and without (fixed $T_{\rm dust} = 110$ K) an AGN component, respectively. We note that results for $T_{\rm dust} = 40$ K with an AGN component are not shown here, as they yield similar or even better agreement with the relation (see Sect.~\ref{Sec: relation}).  The blue square indicates the expected $L_{\rm \cii}$ location of ID9094 if it is an SFG without an AGN. Grey and black solid lines, along with their shaded regions, represent the $L_{\rm \cii}$–SFR relations and $1\sigma$ uncertainties for low-redshift galaxies \citep{delooze2014} and high-redshift ALPINE galaxies \citep[$z\sim4-6$;][]{schaerer2020}, respectively. The orange dashed line shows the relation for low-$z$ composite galaxies and AGNs \citep{delooze2014}. Previous observations of $z>6$ galaxies are plotted as grey points or downward triangles representing 3$\sigma$ upper limits \citep{harikane2020, Schouws2023, Fudamoto2024b}.}
         \label{relation}
\end{figure}

For ID9094, when no AGN component is included in the SED fitting, the SFR derived corresponds to $T_{\rm dust} = 110$ K. Combining this SFR with the 3$\sigma$ upper limits of the $L_{\rm \cii}$ (log($L_{\rm \cii}/L_\odot) < 8.23$; Sect.~\ref{Sec: no_detection}) places ID9094 significantly below the $L_{\rm \cii}$–SFR$_{\rm tot}$ relation, with a deviation of $\sim1$ dex (see Fig.~\ref{relation}). This large offset suggests that the observed properties of ID9094 may not be fully explained by typical star formation processes in dusty star-forming galaxies. In contrast, when including an AGN component in the SED fitting, the resulting SFR with the $T_{\rm dust} = 60$ K template is about four times lower than that returned by SED fitting without an AGN component. In this case, ID9094 lies within the scatter of the $L_{\rm \cii}$–SFR$_{\rm tot}$ relation, indicating consistency with the scaling relation. We note that including an AGN component with $T_{\rm dust} = 40$ K results in a lower SFR and brings ID9094 into even better agreement with the $L_{\rm [CII]}$–SFR$_{\rm tot}$ relation.  

For ID2756, its 3$\sigma$ upper limits of the \cii\, emission line luminosity is log($L_{\rm \cii}/L_\odot) < 8.58$ (Sect.~\ref{Sec: no_detection}). Due to the shallower NOEMA observations, we cannot constrain SFR well with and without AGN components in the SED (as described in Sect.~\ref{Sec: sed}). Therefore, we are unable to place meaningful constraints on its position in the $L_{\rm \cii}$ versus SFR$_{\rm tot}$ plane without deeper FIR observations.

\subsubsection{Caveats} \label{Sec:caveats}

Our analysis provides constraints on the dust and gas properties of the two LRDs. For ID9094, the two independent analyses of $T_{\rm dust}$ and the $L_{\rm [CII]}$–SFR$_{\rm tot}$ relation show notable deviations from typical dusty star-forming galaxies. These distinct properties suggest that additional radiation mechanisms, such as AGN activity, may contribute to the total emission of this source.

However, we emphasize that the interpretations above rely on several assumptions and modeling choices. First, the lower limit on the dust temperature ($T_{\rm dust} \,\gtrsim\, 110$\,K) inferred in the absence of an AGN component is derived under fixed SED parameters and based solely on upper limits in the FIR. The resulting value is therefore model-dependent and sensitive to assumptions such as the dust emissivity index and IR SED shape. Moreover, high dust temperatures alone are not sufficient to definitively rule out a purely star-forming scenario. Although rare, some high-redshift galaxies without clear AGN signatures have exhibited elevated $T_{\rm dust}$ values. For instance, MACS0416-Y1 at $z = 8.31$ has a $T_{\rm dust} \sim 80-116$ K, potentially due to intense starburst activity in a merger-driven system without requiring AGN heating \citep[e.g.,][]{bakx2020, Fudamoto2023, Harshan2024, Sommovigo2022}. Similarly, A2744-YD4 at $z = 8.38$ shows $T_{\rm dust}\sim55-107$ K, which may be attributed to its low metallicity and high star formation rate \citep[e.g.,][]{laporte2017a, behrens2018, Sommovigo2022}. These cases are considered extreme and likely involve compact starbursts and low dust content.

Second, the observed offset from the empirical $L_{\rm [CII]}$–SFR relation is not conclusive evidence against the DSFG scenario. The total SFR estimates are subject to uncertainties, especially when derived from upper limits, and the comparison sample includes a mix of UV-bright and IR-selected galaxies with diverse properties. Moreover, the $L_{\rm [CII]}$–SFR relation exhibits large scatter at high redshift, and many star-forming galaxies also lie below the relation.

In summary, while our results are more consistent with an AGN-driven scenario, the possibility of dusty star formation remains viable given the current data limitations. Future deeper FIR observations will be essential to distinguish between these scenarios.

\section{Impact of LRDs on our understanding of DSFGs}\label{Sec: discussion}

As introduced earlier, the two LRDs analyzed in this study were initially selected as the most massive DSFG candidates at $z_{\rm spec}>7$ in the JWST FRESCO GOODS-North field \citep{Xiao2024}. This misclassification arose due to the overlap in selection criteria between LRDs and DSFGs, particularly for optical/NIR-selected DSFGs -- often referred to as optically dark/faint galaxies \citep[also known as HST-dark galaxies or H-dropouts; e.g.,][]{wangtao2019, Alcalde2019, Williams2019, Xiao2023, Gomez-Guijarro2022a, Barrufet2023, Akins2023, Perez-Gonzalez2023_hstdark, Barro2024}. Typically, DSFGs identified in optical/NIR observations are selected based on their red colors and rest-frame optical faintness, as these characteristics correlate strongly with dust attenuation at similar redshifts \cite[e.g.,][]{wang2016, Xiao2023}. Additionally, the optical faintness criterion helps exclude quiescent galaxies, which are generally optically bright. For instance, DSFGs are often selected using a color cut such as F150W–F444W $> 1.5$ mag and F150W $\gtrsim 26.5$ mag, though specific band choices and thresholds slightly vary across studies.

With the increasing availability of JWST spectroscopic data, emission-line detections have become an additional criterion for DSFG selection \citep[e.g.,][]{Xiao2024, Barrufet2025}. In \citet{Xiao2024}, DSFG candidates were selected based on their red color (F182M $-$ F444W > 1.5 mag) and strong emission lines (e.g., H${\alpha}$+[NII]+[SII] lines or \oi$\lambda\lambda4960,5008$+H${\beta}$ lines). This approach also allows for the identification of LRDs among DSFG candidates. Among 26 DSFG candidates with detected H${\alpha}$ lines ($>8\sigma$), seven ($\sim27\%$) were found to be LRDs, exhibiting broad H$\alpha$ emission lines \citep[see][]{Matthee2024}.

Over the past decade, numerous studies have shown that optically-dark/faint galaxies tend to have high stellar masses and may dominate the massive end of the stellar mass function \citep[e.g.,][]{wangtao2019, Gottumukkala2024}. Additionally, these galaxies could contribute significantly to the cosmic star formation rate density (SFRD) at $z > 3$, with contributions of $\sim10-50\%$ of the SFRD, depending on the methodology (see a direct comparison in \citealt{Xiao2023}; e.g., \citealt{wangtao2019, Williams2019, Gruppioni2020, Fudamoto2021, Talia2021, Enia2022, Shu2022, Barrufet2023, vanderVlugt2023}). Notably, while different studies employ varying selection criteria for optically dark/faint galaxies, some adopt stricter definitions, requiring FIR/submm (e.g., ALMA/NOEMA) detection and/or non-detection in HST. Given that LRDs have been found to 1) lack FIR/submm detections \citep[see also e.g.,][]{Labbe2025, Williams2024_RED}, 2) exhibit a characteristic ``v-shaped" SED (indicating detection in HST), they are unlikely to contaminate the DSFG sample in these studies.

On a global scale, these findings suggest that LRD contamination should not significantly bias SFRD estimates derived from DSFG samples. However, their presence may still impact number density estimates of optical/NIR-selected DSFG populations, particularly when broader color-based selection criteria are used. A study with a larger sample is needed to determine how much the number density estimates of DSFG populations are affected by LRDs. These results highlight the importance of refining DSFG selection methods to minimize potential contamination from AGN-dominated LRDs, ensuring a more accurate census of dust-obscured star formation in the early Universe.

\section{Conclusions}\label{Sec: conclusion}

In this paper, we present a study of two LRDs, ID9094 and ID2756, recently discovered in the JWST FRESCO GOODS-North field at $z_{\rm spec}>7$. These sources, characterized by their red, point-source-like morphology and "v-shaped" SEDs, exhibit broad H$\beta$ emission lines, suggesting them as AGN candidates. Using NOEMA observations targeting the \cii$\,158\,{\rm \mu m}$ emission line, and the 1.3 mm dust continuum, we obtain the following key results:

\begin{itemize}

    \item Both sources are undetected in the \cii\ and 1.3 mm continuum. For ID9094, the $3\sigma$ upper limits on \cii\ and 1.3 mm continuum are far below expectations for a typical DSFG, where it would have been detected at $>16\sigma$ in \cii\ and $>3\sigma$ at 1.3 mm. For ID2756, the expected $>5\sigma$ detection in \cii\ is also not observed.
    \item Benefiting from the 1.3 mm upper limits, we perform UV-to-FIR SED fitting, which provides the first hint of an AGN scenario in this work. For ID9094, without including an AGN component, the SED fit requires an abnormally high dust temperature ($T_{\rm dust} \,\gtrsim\, 110$ K), significantly above typical values for SFGs at similar redshifts. Including an AGN component yields more physically plausible fits with dust temperatures that can be as low as $\sim$40–60 K. However, we caution that the minimum $T_{\rm dust}$ is strongly linked to the AGN fraction contributing to the total IR luminosity -- higher AGN fractions allow for lower $T_{\rm dust}$ values. For ID2756, the SED fitting remains inconclusive due to shallower NOEMA data, highlighting the need for deeper FIR/submillimeter observations.  
    \item Using the \cii\ upper limits, we compare the sources to the empirical $L_{\rm \cii}$–SFR$_{\rm tot}$ relation, providing a second hint of an AGN scenario in this work. For ID9094, its \cii\ upper limit places it significantly below the scaling relation when no AGN component is included, but consistent with the relation when an AGN contribution is considered. The case for ID2756 remains uncertain due to limited NOEMA depth.
\end{itemize}

These results highlight the value of FIR constraints, when combined with shorter-wavelength photometry, in revealing the nature of LRDs -- particularly when spectroscopic data are unavailable.  For ID9094, the results are consistent with a scenario in which AGN activity contributes to the observed emission, though a dusty star-forming origin cannot be fully excluded given the data limitations. We also note that both sources are detected in JWST/MIRI F1000W, and ID9094 is additionally detected in F2100W, supporting the presence of hot dust emission (G. Leung, priv. comm.). For ID2756, the current observations are insufficient to determine the dominant energy source. We emphasize that our interpretation remains model-dependent and subject to assumptions in the SED fitting and limited FIR constraints. More broadly, this study demonstrates the critical role of FIR data in clarifying the nature of LRDs. Future deeper FIR/submillimeter campaigns, along with high-resolution spectroscopy and multi-wavelength data, will be essential to fully understand the physical properties of these LRDs and their role in early galaxy evolution.

\begin{acknowledgements}
We are very grateful to the anonymous referee for instructive comments, which helped improve the overall quality and strengthen the analysis of this work. We thank Andrea Weibel for assistance with the HST and JWST photometric measurements used in this paper.
This work is based on observations carried out under project number S23CY with the IRAM NOEMA Interferometer. IRAM is supported by INSU/CNRS (France), MPG (Germany) and IGN (Spain). 
This work is based in part on observations made with the NASA/ESA/CSA James Webb Space Telescope. The data were obtained from the Mikulski Archive for Space Telescopes at the Space Telescope Science Institute, which is operated by the Association of Universities for Research in Astronomy, Inc., under NASA contract NAS 5-03127 for JWST. These observations are associated with programs \#1895 and \#4762. 
Support for programs \#1895 and \#4762 was provided by NASA through a grant from the Space Telescope Science Institute, which is operated by the Association of Universities for Research in Astronomy, Inc., under NASA contract NAS 5-03127.

This work has received funding from the Swiss State Secretariat for Education, Research and Innovation (SERI) under contract number MB22.00072, as well as from the Swiss National Science Foundation (SNSF) through project grant 200020\_207349. 
The Cosmic Dawn Center (DAWN) is funded by the Danish National Research Foundation under grant DNRF140.
This work is partially supported from the National Natural Science Foundation of China (12073003, 11721303, 11991052), and the China Manned Space Project (CMS-CSST-2021-A04 and CMS-CSST-2021-A06). 
Y.F.\ is supported by JSPS KAKENHI Grant Numbers JP22K21349 and JP23K13149.
M.V.\ gratefully acknowledges financial support from the Independent Research Fund Denmark via grant numbers DFF 8021-00130 and  3103-00146 and from the Carlsberg Foundation via grant CF23-0417.
VK acknowledges support from the University of Texas at Austin Cosmic Frontier Center.
S.F. acknowledges support from NASA through the NASA Hubble Fellowship grant HST-HF2-51505.001-A awarded by the Space Telescope Science Institute, which is operated by the Association of Universities for Research in Astronomy, Incorporated, under NASA contract NAS5-26555.
Support for this work for RPN was provided by NASA through the NASA Hubble Fellowship grant HST-HF2-51515.001-A awarded by the Space Telescope Science Institute, which is operated by the Association of Universities for Research in Astronomy, Incorporated, under NASA contract NAS5-26555.

\end{acknowledgements}

\bibliography{reference_update}{}
\bibliographystyle{aa}

\onecolumn
\begin{appendix}

\section{Best-fit SED of ID2756 with $T_{\rm dust}=40$ K.}\label{appendix1} 

\begin{figure*}[h!]
\centering
\includegraphics[width=15cm]{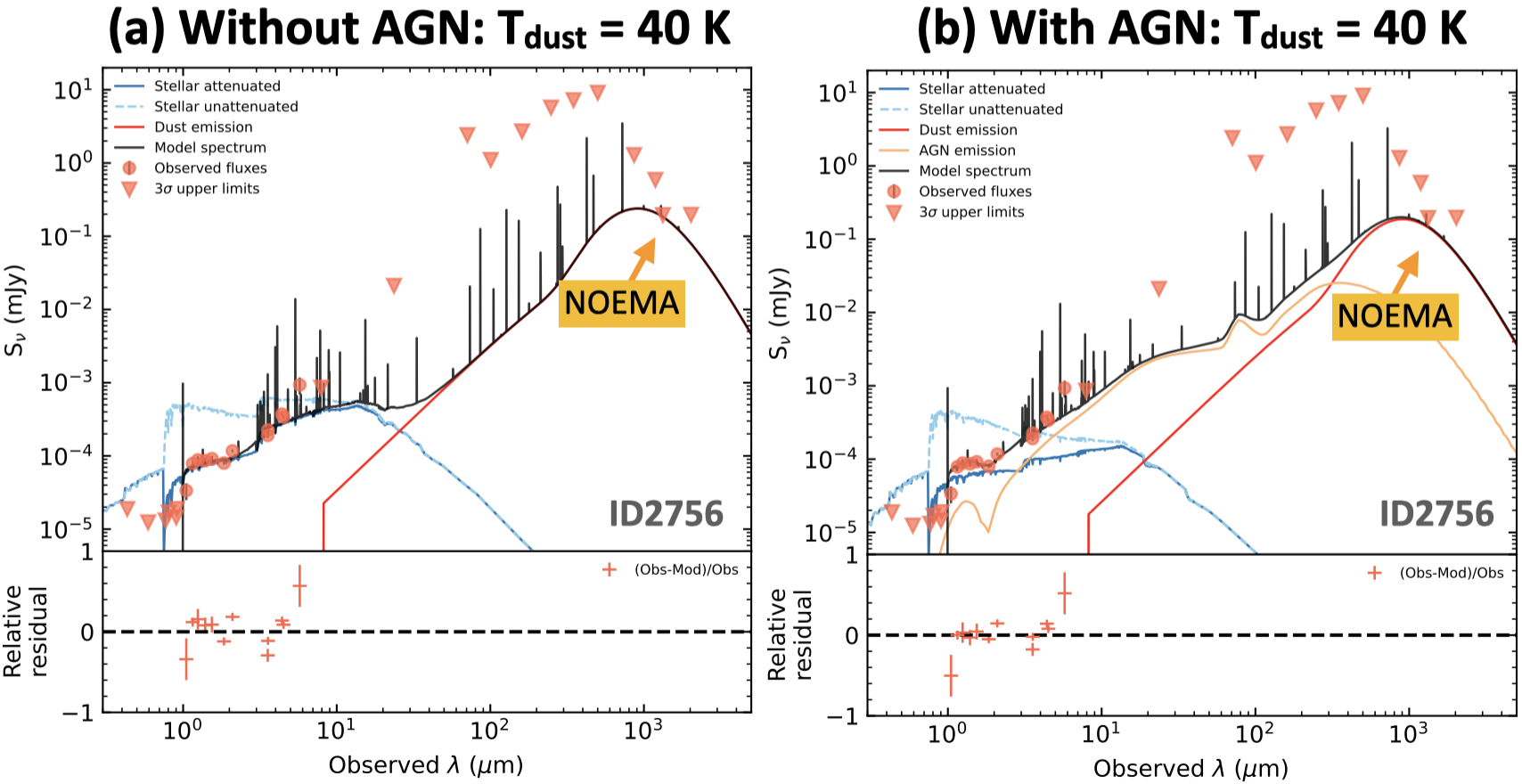}
\caption{Similar to Fig.~\ref{9094_sed}, but for ID2756. The $3\sigma$ upper limit at 1.3 mm is not sufficiently deep to rule out the low $T_{\rm dust}$ solution, and also rule out the scenario without an AGN. Deeper FIR/submm observations, at least similar to those for ID9094 (rms level $\sim$30.4 $\mu$Jy beam$^{-1}$ at 1.3 mm), are needed. 
      }
         \label{2756_sed_40k}
\end{figure*}

\end{appendix}

\end{document}